\documentclass[runningheads]{llncs}

\usepackage{multirow}
\usepackage{booktabs}
 
\usepackage{eccv}

\usepackage{eccvabbrv} 

\usepackage{graphicx}
\usepackage{booktabs}

\usepackage[accsupp]{axessibility}  

\usepackage{hyperref}

\usepackage{orcidlink}

\begin{document}

\title{Mesh2GS: White-Box 3DGS Construction via Plenoptic Sampling} 

\titlerunning{Abbreviated paper title}

\author{ Haoran Zhu\inst{1}\orcidlink{0009-0002-3705-0768} \and
	Huangsheng Du\inst{1}\orcidlink{0009-0001-0987-3735} \and
	 Youcheng Cai\inst{1}\orcidlink{0000-0002-0599-8418} \and
	 Jingyang Meng\inst{1}\orcidlink{0009-0000-6494-0868} \and
	 Ligang Liu\inst{1}\orcidlink{0000-0003-4352-1431}
}

\authorrunning{F.~Author et al.}

\institute{University of Science and Technology of China}
\maketitle

\begin{abstract}
3D Gaussian Splatting (3DGS) has emerged as a promising method for high-quality, real-time 3D reconstruction. To associate 3DGS with mesh representations, existing methods primarily focus on 3DGS-to-mesh reconstruction from multi-view images. In contrast, the problem of converting a mesh into 3DGS has received comparatively less attention. Instead of relying on heuristic strategies that bind 3D Gaussians to the mesh, we propose a novel white-box 3DGS construction framework, termed Mesh2GS, which generates 3DGS directly from mesh geometry based on plenoptic sampling theory, achieving Nyquist-level performance for high-quality global illumination rendering. Firstly, we propose a plenoptic sampling guided 3DGS construction strategy that theoretically derives the minimum sampling rate of the sampled views and the distribution of 3D Gaussians. Second, we propose a novel 3DGS update procedure with albedo--shading decomposition for efficient global-illumination capture. Finally, we introduce a neural illumination enhancement module to handle non-Lambertian effects. Experimental results demonstrate that our method surpasses state-of-the-art baselines and is practically effective for both real-time shared rendering and non-Lambertian effects capturing specular highlights. The project code will be released upon acceptance.
\end{abstract}

\section{Introduction}

Rendering realistic depictions of real-world environments has long been a significant challenge, with numerous practical applications in fields such as computer vision and computer graphics. Most graphics and game engines \cite{karis2013real} employ polygonal mesh-based representations of scenes and objects owing to their intuitive structure and ease of editing. Although photorealistic rendering is achievable, computing global illumination typically requires expensive Monte Carlo path tracing computations \cite{RN2,RN3} .

3D Gaussian Splatting (3DGS) \cite{RN13} has attracted considerable interest for its high-quality, real-time novel view synthesis. 3DGS excels at capturing intricate scene details with high fidelity due to their flexibility in splat positions and shapes. 3DGS naturally serve as an effective caching representation, capable of caching the global illumination of a scene, thereby avoiding complex path tracing computations and enabling efficient rendering, thus holding significant potential for applications such as AR, VR, and shared rendering. 

Existing methods \cite{huang20242d, RN55} have focused on reconstructing a 3D mesh from 3DGS using given multi-view images, despite the unknown structure of the scene—a process we refer to as “black-box” 3DGS construction. Despite extensive efforts on 3DGS-to-mesh reconstruction, the problem of converting a mesh to 3DGS has also recently attracted increasing attention. Given the mesh geometry of a scene, conversion to a 3DGS representation enables the baking of global illumination, which can be efficiently utilized in real-time rendering applications for multiple users from arbitrary viewpoints. Recently, a common approach involves tightly binding 3D Gaussian splats to mesh triangles \cite{RN55, RN56, choi2024meshgs, waczynska2024games}. SuGaR \cite{RN55} captures fine object-level structures and tightly binds the 3D Gaussian splats to the surface of the mesh. GaussianMesh \cite{RN56} introduces a mesh-based GS representation that learns a Gaussian distribution coupled with an explicit mesh to enable deformations. Although these heuristic methods are intuitive, they lack a rigorous formulation of the view sampling and Gaussian distribution requirements. When faced with a new scene, users of these methods are limited to trial-and-error to determine whether a set of sampled views and 3D Gaussians can cache realistic global illumination in the environment.

To tackle the aforementioned limitations, we propose a new problem termed "white-box" 3DGS construction, which aims to determine the distribution of 3D Gaussians and the density of sampled views based on the underlying mesh geometry of the scene, grounded within the plenoptic sampling theory \cite{RN14}. However, this problem presents three key challenges: (1) accurately specifying how densely sampled views should cover the scene; (2) designing the 3D Gaussian distribution to enable reliable caching of global illumination; and (3) handling non-Lambertian effects that are difficult for vanilla 3DGS.

To address the challenges above, we propose a novel white-box 3DGS construction framework, referred to as Mesh2GS, which constructs 3DGS from the mesh geometry of a scene, achieving Nyquist-level performance for high-quality global illumination rendering. Firstly, we propose a plenoptic sampling guided 3DGS construction strategy, which theoretically derives the minimum sampling rate of the sampled views and the distribution of 3D Gaussians that enables effective caching of global illumination effects. Secondly, we introduce a novel 3DGS update procedure with an albedo--shading decomposition strategy to efficiently capture global illumination. We extensively validate our derived view sampling and Gaussian distribution requirements and demonstrate that our approach qualitatively and quantitatively outperforms heuristic 3DGS construction methods. Finally, we introduce Neural Illumination Enhancement (NIE) module applied only to non-Lambertian regions to model view-dependent highlights. 

To demonstrate practical value, we apply our method to real-time shared rendering scenarios, which achieves state-of-the-art performance, enabling high-quality, real-time caching for multi-viewer environments. To the best of our knowledge, our approach is the first framework for white-box 3DGS construction based on plenoptic sampling theory.

To summarize, we provide the following contributions:
\begin{itemize}
    \item We develop a white-box 3DGS framework that constructs 3DGS directly from the mesh geometry, enabling high-quality global illumination rendering.
	\item We theoretically derive the minimum sampling rate for sampled views and the distribution of 3D Gaussians, grounded in the plenoptic sampling theory.
    \item We combine an albedo--shading decomposition strategy with a Neural Illumination Enhancement (NIE), enabling non-Lambertian effects capture. 
\end{itemize}

\section{Related Work}

\subsection{{Global Illumination Caching}}
The problem of global illumination dates back to the formulation of the rendering equation \cite{RN15}, which simulates the interaction between light and matter through a statistical solution based on path tracing, albeit with high computational cost. 
To achieve real-time rendering, an appealing approach is caching global illumination, which can be classified as precomputation caching or real-time caching.

Precomputation caching of Precomputed Radiance Transfer (PRT) is one of the most established techniques. The original paper \cite{RN16} introduces spherical harmonic functions to represent lighting, supporting low-frequency global illumination effects. Subsequent works extend this by enabling self-shadowing \cite{RN17}, near-field lighting \cite{RN18}. Recently, to handle fully dynamic scenes, LightFormer \cite{RN19} encodes the information from every light present in the scene, preserving high-frequency shading details. NeLT \cite{RN8} proposes a modular neural rendering framework that models object-oriented light transfer, supporting flexible rendering and editing. Nevertheless, PRT approaches usually suffer from generalization problems, requiring the scene to be static or known in advance.

Real-time caching techniques utilize online caching mechanisms that enable dynamic updates to caches. Probe-based approaches \cite{RN5,RN20} typically place light probes in world-space grids; these probes are dynamically updated and interpolated to approximate multi-bounce indirect illumination. Furthermore, Wright et al. \cite{RN21} introduce Screen Space Radiance Caching, which places probes directly on the visible surfaces in screen space to reduce the effects of light and occlusion. Recently, Müller et al. \cite{RN9} propose neural radiance caching (NRC), which can learn indirect lighting online and simulate infinite-bounce transport by iterating only a few-bounce training updates. Our framework is similar to NRC, updating during real-time rendering to adapt to changes in scenes and lighting conditions without the need for pre-generated training datasets. 
In contrast, our framework caches global illumination effects based on 3DGS, adapts to real-time multi-viewer rendering, and supports non-Lambertian appearance modeling through Neural Illumination Enhancement module. By comparison, existing methods either cache only partial global-illumination effects, are limited to single-viewer scenarios, or struggle with non-Lambertian effects.

\subsection{3D Gaussian Splatting}
Recently, 3D Gaussian Splatting (3DGS) \cite{RN13} explicitly represents a scene using 3D Gaussians and employs efficient differentiable splatting, thereby achieving high-quality rendering while maintaining efficient rendering speeds. To establish a connection between 3DGS and mesh representations, SuGaR \cite{RN55} proposes directly binding Gaussians to the mesh surface, enabling accurate mesh reconstruction. GaussianMesh \cite{RN56} introduces a new mesh-based GS representation that learns a Gaussian distribution anchored to an explicit mesh, which achieves a large-scale Gaussian deformation. GaMeS \cite{waczynska2024games} tightly binding 3D Gaussian to meshes which parametrizes flat Gaussians by the vertices of the triangle mesh. Recently, MeshGS \cite{choi2024meshgs} proposes a distance-based Gaussian splatting method that aligns 3D Gaussians with the mesh surface through geometric regularization to enhance rendering quality. However, these methods typically construct 3D Gaussians from meshes using heuristic strategies. In contrast, our white-box 3DGS construction strategy is grounded in plenoptic sampling theory, which precisely constrains Gaussian distribution and camera sampling density, enabling reliable global-illumination caching. 

\subsection{{Image-based Rendering}}
Image-based rendering (IBR) is a fundamental problem in computer graphics that reconstructs novel views through light field sample from sampled views. 

Classically, light field rendering \cite{RN27} represents light fields using 2D slices, allowing new views to be rendered by extracting appropriate slices. Lumigraph rendering \cite{RN28} describes a rendering framework based on the plenoptic function without requiring any geometric knowledge. Subsequently, Chai et al. \cite{RN14} study plenoptic sampling to derive the minimum sampling rate for light field rendering, which reflects the relationship between scene depth accuracy and the number of sampled viewpoints. Later, Zhang et al. \cite{RN29} extend this work with surface plenoptic function to generic scenes such as scenes with non-Lambertian and occlusion. Further work takes stereo imagery into consideration for view synthesis. For instance, Mitra et al. \cite{RN30} model the light field based on the disparity pattern of the scene, which depends on the depth of the scene relative to the plane of focus. Zhang et al. \cite{RN31} propose a phase-based method that reconstructs a 4D light field from a micro-baseline stereo pair. 

Recently, researchers have focused on powerful deep learning techniques to improve IBR. DeepStereo \cite{RN32}, learning-based light field camera interpolation \cite{RN33} and learning-based single view light field synthesis \cite{RN34} use convolutional neural networks to estimate scene geometry for novel view synthesis in the light field. To synthesize new views from narrow-baseline stereo pairs, Zhou et al. \cite{RN35} propose a new layered representation of multiplane images (MPIs) to represent geometry and texture. Furthermore, LLFF \cite{RN36} extend the MPIs by combining the plenoptic sampling theory to reconstruct the light field, significantly reducing the number of required sampled images. Similarly, to leverage the significant benefits of prescriptive sampling, we construct 3DGS based on plenoptic sampling theory, enabling efficient caching of global illumination.


\section{Overview}
\begin{figure}[t]
	\includegraphics[width=\linewidth]{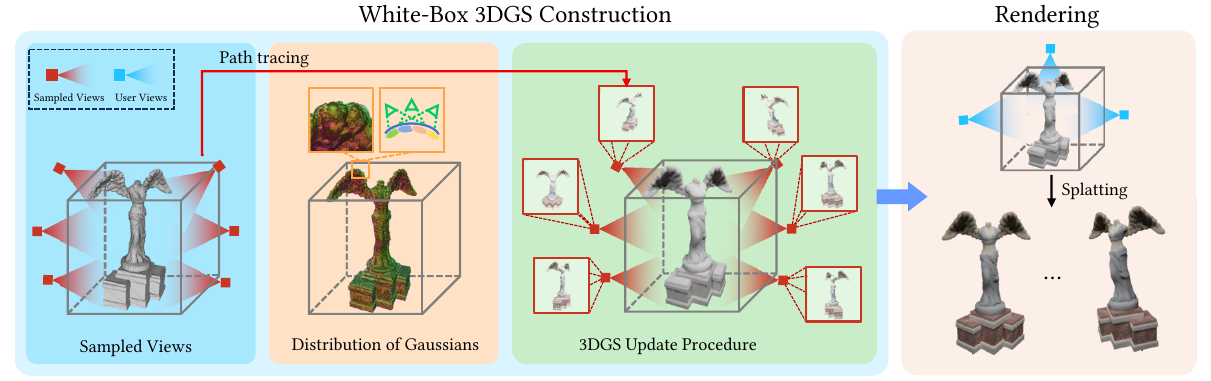}
	\caption{Overview of our white-box 3DGS construction framework. Given scene geometry, we first explicitly construct 3DGS based on plenoptic sampling theory for reliable caching of global illumination effects. Subsequently, the 3DGS update procedure utilizes path-traced renderings to efficiently update 3DGS and capture global illumination.}
	\label{fig1}
\end{figure}

An overview of our framework is shown in Fig.~\ref{fig1}. 
First, we propose a plenoptic sampling guided 3DGS construction strategy that derives the admissible view-sampling density and allocates 3D Gaussians directly from mesh geometry, enabling Nyquist-level global-illumination caching in a white-box manner. 
Then, we introduce an efficient 3DGS update procedure with albedo--shading decomposition, together with NIE strategy for modeling non-Lambertian effects.

\subsection{{Preliminary}}

\subsubsection{3D Gaussian Splatting.}
3DGS \cite{RN13} employs anisotropic 3D Gaussians as primitives to represent 3D scenes and utilizes efficient splatting to achieve fast rendering and optimization. Specifically, each Gaussian can be expressed as,
\begin{equation}
G(\mathbf{x})=e^{-\frac1 2 (\mathbf{x}-\mathbf{p})^T\Sigma^{-1}(\mathbf{x}-\mathbf{p})}
\end{equation}
where $\mathbf{p}$ represents the position of the Gaussian, and $\Sigma = R S S^{T} R^{T}$ denotes the covariance matrix, which is factorized into a scaling matrix $S$ and a rotation matrix $R$. Additionally, the Gaussian includes spherical harmonics (SH) coefficients to represent anisotropic colors.

For rendering, 3D Gaussians are transformed into camera space using an approximate affine transformation and are subsequently projected onto the 2D image plane in depth-sorted order. The color for each pixel is then calculated by using a forward alpha-blending, 
\begin{equation}
C=\sum_{i\in N}c_i\alpha_i\prod_{j=1}^{i-1}(1-\alpha_j)
\end{equation}
where $\alpha_i$ denotes the projected contribution of the $ith$ Gaussian on the 2D image plane, scaled by its opacity, while $c_i$ denotes its color computed from the view direction and the corresponding SH coefficients.

\subsubsection{Plenoptic Sampling Theory.} 
The initial work on plenoptic sampling \cite{RN14} demonstrates that the light field of any scene with a depth between the minimum $z_{min}$ and the $z_{max}$ has its continuous Fourier spectral support bounded in the frequency domain, and derives the maximum camera sampling interval for light field rendering. Assuming Lambertian scenes without occlusion and uniform sampling of the light field, the spectral support of the light field takes the form of a double-wedge shape. Therefore, according to the Nyquist sampling theorem, the maximum camera sampling interval $\Delta t$ for a light field is, 
\begin{equation}
\Delta t = \frac{1}{K_{v}f(1/z_{min}-1/z_{max})}\label{con:sampling}
\end{equation}
where $f$ is the camera focal length, $K_{v}$ is the highest spatial frequency represented in the sampled light field, which can be computed as,
\begin{equation}
K_v = min(B_v,\frac{1}{2\Delta v})\label{con:max frequency}
\end{equation}
where $B_v$ represents the highest spatial frequency in the continuous light field, and $\Delta v$ is the pixel size of the camera. From an intuitive perspective, the above formulation indicates that the maximum camera sampling interval is equivalent to the condition where the maximum disparity is less than the smallest resolvable feature on the image plane.

\subsection{Problem Statement}

Given scene geometry $G$, the sampling cameras with focal length $f$, pixel size $\Delta v$ (image space sampling interval), field of view (FOV) $\theta$, the observable depth range $[z_{min}, z_{max}]$, we now state our problem of white-box 3DGS construction.

Our goal is to reconstruct the full-scene light field while minimizing the number of Gaussians $n_g$ and sampled views $n_c$, which corresponds to determining the maximum camera sampling interval $\Delta t$ and constructing the minimal set of Gaussians that achieves Nyquist-level rendering performance. 

\section{Plenoptic Sampling Guided 3DGS Construction}
To construct 3DGS from a given scene, a straightforward strategy is to render a large number of multi-view images using path tracing and directly optimize 3DGS from these observations. However, accurately capturing global illumination typically requires dense viewpoint sampling, which introduces substantial path tracing costs and leads to numerous redundant observations. Consequently, the resulting 3DGS representation often contains a large number of redundant Gaussians and incurs unnecessary optimization overhead. In fact, prior studies have shown that a significant portion of Gaussians can be pruned with only minor degradation in rendering quality, indicating considerable redundancy in existing 3DGS constructions. Existing methods \cite{RN55, RN56} typically bind 3D Gaussians to mesh triangles to achieve realistic deformation of 3DGS. However, such approaches are not suitable for caching global illumination.

Therefore, we propose a white-box 3DGS construction strategy guided by plenoptic sampling theory \cite{RN14}. Our approach explicitly constructs 3D Gaussians based on theoretical sampling principles, enabling efficient 3DGS construction while supporting global illumination caching. Although the 3D Gaussians are explicitly initialized based on the plenoptic sampling theory, purely explicit Gaussian representations remain limited in modeling non-Lambertian effects. To address this issue, we introduce a Neural Illumination Enhancement module that is only activated for regions exhibiting complex appearance variations.

From a signal processing perspective, 3D Gaussians can be viewed as a set of spatial sampling points distributed in world space. The splatting operation can be regarded as a low-pass filter that reconstructs discrete signals in world space into continuous signals and resamples them on the image plane. Therefore, the construction of 3D Gaussians, which inherently reconstructs the light field, is closely tied to the configuration of the sampling cameras. According to plenoptic sampling theory, achieving Nyquist-level light field reconstruction depends on factors such as the camera sampling interval $\Delta t$, pixel size $\Delta v$, and the scene geometry, particularly the minimum observation distance $z_{min}$. These factors also place a lower bound on the spatial sampling density of 3D Gaussians in world space, since higher-frequency details cannot be captured by the cameras. Consequently, by setting the maximum camera sampling interval and the minimum spatial sampling density of 3D Gaussians, one can jointly minimize the number of sampled views $n_c$ and the number of Gaussians $n_g$, while still achieving Nyquist-level reconstruction.


\begin{figure}[t]
	\centering
	\includegraphics[width=0.25\linewidth]{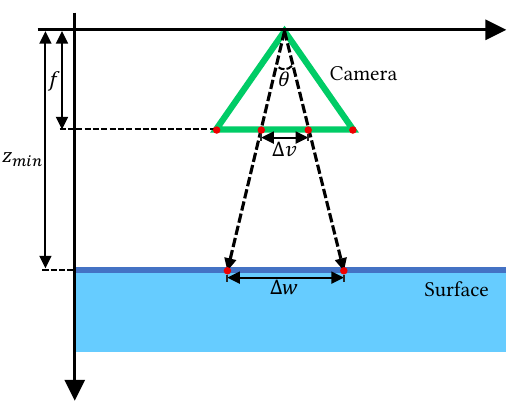}
    \includegraphics[width=0.25\linewidth]{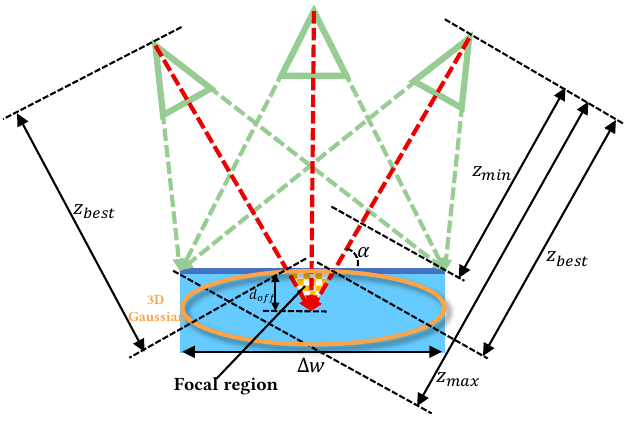}
    \includegraphics[width=0.4\linewidth]{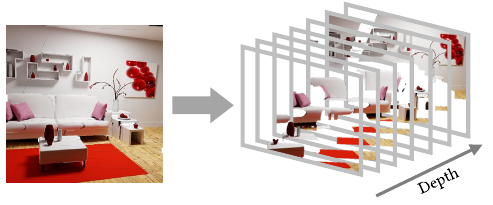}
	\caption{Left: Sampling interval of 3D Gaussians $\Delta w$ in world space based. Mid: 3D Gaussians should be distributed just below the visible surfaces of scenes. Right: The 3DGS can naturally represent numerous depth ranges, rendering images at different depth layers.
		on the sampling interval $\Delta v$.
	}
	\label{fig2}
\end{figure}

\subsection{Sampling Interval of Camera Views }
Plenoptic sampling theory \cite{RN14} further demonstrates that the sampling rate can be reduced with more knowledge of scene geometry. By decomposing a scene into $D$ depth ranges and sampling the light field independently within each range, the maximum camera sampling interval can be increased by a factor of $D$, 

\begin{equation}
\Delta t = \frac{D}{K_{v}f(1/z_{min}-1/z_{max})}\label{con:sampling D}
\end{equation}
This suggests that when the number of depth layers is sufficient, a single viewpoint is adequate to reconstruct the light field within its visible range, as also demonstrated by \cite{RN36}. The 3DGS can naturally represent numerous depth ranges, rendering images at different depth layers, as shown in Fig. \ref{fig2}, which significantly increases the maximum camera sampling interval. Therefore, we assert that the light field of a scene can be reconstructed by $n_c$ sampled views that covers all visible surfaces in the scene with the minimum observation distance of $z_{min}$. Specifically, we uniformly sample views from a spherical space around each object in the scene, maintaining a minimum observation distance of $z_{min}$ while ensuring adequate coverage of all visible surfaces.



\subsection{Distribution of 3D Gaussians}
\textbf{Sampling Interval of 3D Gaussians.}
As shown in Fig. \ref{fig2}, we derive the sampling interval $\Delta w$ in world space based on the sampling interval $\Delta v$ in image space, which can be computed as, 
\begin{equation}
\Delta w = \frac{z_{min}\Delta v}{f},\Delta v = \frac{2f \tan\frac{\theta}{2}}{R}
\end{equation}
where $\theta$  is the angle based on the sampling interval  $\Delta v$, and $R\in(H,W)$ is the resolution of the camera. We set the depth as $z_{min}$ to satisfy the required sampling frequency.

\textbf{The Position of 3D Gaussians.}
Most existing works \cite{RN56,RN57,RN58} suggest that 3D Gaussians are placed on the surface of the scene, strictly aligning with the triangles. However, we observe that 3D Gaussians should be distributed just below the visible surfaces of scenes. Specifically, Plenoptic sampling theory \cite{RN14} defines the optimal depth $\frac{1}{z_{best}} =\frac{1}{2}(\frac{1}{z_{min}} +\frac{1}{z_{max}} )$, which yields the best rendering quality. Given a small region with the sampling interval $\Delta w$ and the minimum observation distance $z_{min}$, we consider the intersections of the optimal depth with the surface, as well as the intersections of the principal axes of all cameras. A focal region is thereby obtained (as shown in the yellow region of Fig. \ref{fig2}), which is distributed beneath the surface. In this way, the downward offset of the surface of the focal region can be computed as follows,
\begin{equation}
d_{off} = \frac{\Delta w^2 sin\alpha}{2(z_{min}+z_{max})}
\end{equation}
where $\alpha$ is the angle between the surface and the principal axis of the camera, and $d_{off} \in[0,\frac {\Delta w^2}{4z_{min}}]$. Consequently, we set the Gaussians to be positioned at a downward offset of the surface by  $\frac {\Delta w^2}{4z_{min}}$ , aligning them with the surface.

\textbf{Initialization of Attributes for 3D Gaussians.}
(1) the number of Gaussians: we uniformly sample the points on the mesh according to the sampling interval $\Delta w$; (2) the scale of Gaussians: we set $r_x=r_y=\Delta w$, and $r_z=\frac {\Delta w^2}{4z_{min}}$; (3) the position of Gaussians: the Gaussians are positioned at a downward offset of the surface by $\frac {\Delta w^2}{4z_{min}}$; (4) the rotation of Gaussians: the Gaussians are aligned with the surface; (5) the opacity of Gaussians: we initialize all Gaussians with an opacity of $0.5$




\subsection{Neural Illumination Enhancement}

Although 3D Gaussian Splatting (3DGS) efficiently represents scene geometry and base appearance, its explicit Gaussian representation is limited in modeling complex non-Lambertian effects. In particular, view-dependent effects such as specular highlights exhibit strong directional variations that are difficult to capture using explicit Gaussians alone.

To address this limitation, we introduce a Neural Illumination Enhancement module that complements the explicit 3DGS representation. The module is selectively applied to non-Lambertian effects, while Lambertian effects are rendered using the standard 3DGS pipeline, preserving the efficiency of explicit splatting.

Specifically, we employ a lightweight Multilayer Perceptron (MLP) to predict a view-dependent color residual. The network takes as input the spatial position, reflection direction, and several material-related features. Different encodings are applied to different inputs: spatial positions are encoded using a HashGrid encoding to capture local spatial structures, reflection directions are encoded using Spherical Harmonics (SH) to model directional effects, while other features—including view direction, surface normals, Fresnel term, and roughness—are directly used as identity features.

The final rendered image is computed as
\begin{equation}
I = I_g + I_n,
\end{equation}
where $I_g$ denotes the gaussian image rendered by the explicit 3DGS representation, and $I_n$ represents the residual predicted by the neural module. Since the network only models view-dependent residuals, most regions remain efficiently represented by explicit Gaussians.

\begin{figure}[t]
	\centering
	\includegraphics[width = 0.75\linewidth]{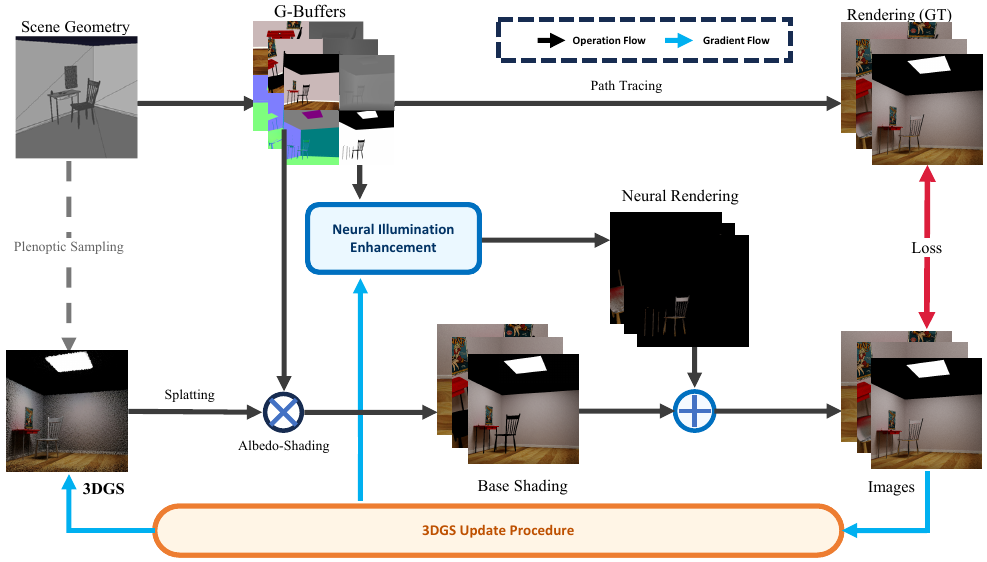}
	\caption{Our 3DGS update procedure employs path-traced renderings from the sampled views to optimize the color attributes of the 3D Gaussians. The albedo-shading decomposition strategy is employed to improve performance.
	}
	\label{fig9}
\end{figure}

\section{3DGS Update Procedure}
\label{Sec5}
To efficiently optimize 3DGS for caching global illumination, our 3DGS update procedure involves: (1) path-traced renderings using sampled views; (2) updating only the color attributes; (3) an albedo-shading decomposition strategy; and (4) neural illumination enhancement module, as illustrated in Fig. \ref{fig9}.

Specifically, after constructing the 3DGS, all attributes of the 3D Gaussians, except for color values, are frozen to enable efficient caching of global illumination in dynamic scenes, in which RGB values are used instead of spherical harmonics (SH) coefficients to represent radiance in Lambertian scenes. Instead of directly optimizing 3DGS to fit path-traced renderings, we decompose the albedo and shading of the scene and guide the 3DGS to cache the shading component. This approach offers two advantages: (1) it lowers the frequency of the reconstructed light field, thereby reducing the required number of 3D Gaussians; and (2) it improves optimization stability and leads to faster convergence. Specifically, the albedo image $I_a$ is obtained from G-Buffers and the shading image $I_s$ is splatted by the gaussians. The gaussian image is represented as $I_g=I_a*I_s$.

Beyond scene reconstruction tasks, if the update rate is sufficiently high, 3D Gaussians can also support more challenging real-time shared rendering applications. In this setting, we can sample only a subset of viewpoints and update the color attributes of the 3D Gaussians in real time, enabling fast rendering from an arbitrary number of views at arbitrary positions. To achieve real-time performance, we optimize the 3DGS using noisy 1-spp path-traced renderings with an exponential moving average (EMA) strategy \cite{RN9}. Our framework adopts a multi-GPU optimization setting. Additional implementation details of our experimental setup can be found in the supplementary material.

\section{Experiments}
\subsection{Dataset and Metric}

We evaluate our method on 20 scenes, including indoor environments, outdoor scenes, and object-level scenes, covering a wide range of geometric structures, lighting conditions, and material properties. For each scene, we construct the corresponding 3D Gaussian representation from sampled viewpoints. The sampling configuration and initialization parameters are automatically determined based on the scene geometry, allowing our approach to adapt to scenes with different scales and complexities. 

All sampling cameras share the same resolution and use a field of view (FOV) of $45^{\circ}$. For quantitative evaluation, we adopt PSNR, SSIM \cite{RN64}, LPIPS \cite{zhang2018unreasonable}, and FLIP \cite{RN65} to measure the perceptual quality of the rendered results.

\subsection{Implementation Details}

We implement our framework in CUDA. All scenes are created in the Mitsuba renderer, which is also used to generate path-traced supervision images. All experiments are conducted on a single NVIDIA RTX 4090 GPU.

\subsection{Comparisons and Evaluations}

We evaluate the performance of the proposed method through both qualitative and quantitative comparisons with several state-of-the-art approaches, including Neural Radiance Caching (NRC)\cite{RN9}, vanilla 3DGS \cite{RN13}, SuGaR \cite{RN55}, GaussianMesh \cite{RN56}, and GaMeS \cite{waczynska2024games}. Our primary focus is on the ability of the constructed 3D Gaussians to cache global illumination within the scene.

Vanilla 3DGS \cite{RN13} serves as the baseline for novel view synthesis by reconstructing a 3D Gaussian representation from sampled views. We further compare with several extensions based on Gaussian representations, including SuGaR, GaMeS, and GaussianMesh. These approaches build upon 3DGS by introducing additional structural constraints or geometric modeling strategies to improve geometric consistency and surface representation. 

In addition, we compare our approach with NRC, which learns the radiance distribution in path tracing using a neural network to cache and predict lighting information. In this section, NRC is primarily treated as a neural caching method to evaluate different approaches in terms of illumination modeling and reflection representation. All viewpoints are uniformly sampled within the scene and rendered using path tracing with 4096 samples per pixel (spp).

In our experiments, the proposed method consistently outperforms the comparison methods. Table~\ref{table1} reports the quantitative results, showing that our approach achieves better performance than the four baselines in terms of PSNR, SSIM, and LPIPS.

\begin{figure*}[t]
	\centering
	\includegraphics[width=\linewidth]{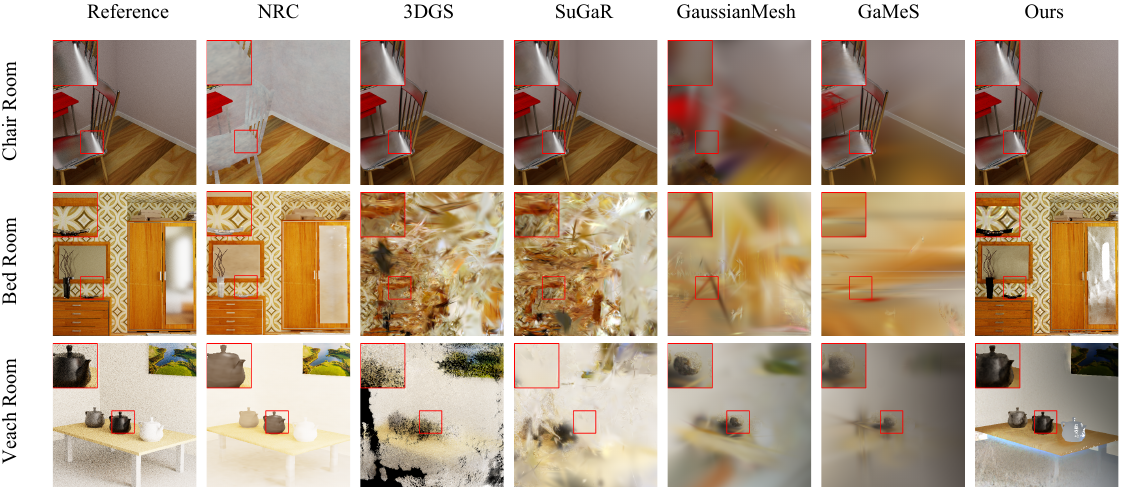}
	\caption{Qualitative comparisons of our approach with state-of-the-art methods. Our method produces high-quality rendering results. Please zoom in for a better visualization. An accompanying video is provided for dynamic qualitative comparisons.
	}
	\label{fig13}
\end{figure*}

The results demonstrate that our method is capable of caching complex global illumination while preserving fine texture details and capturing specular highlights and complex reflection effects from non-Lambertian effects. Qualitative comparisons are shown in Fig.~\ref{fig13}, where our method produces high-quality renderings, particularly in preserving high-frequency details. Additional details are provided in the supplementary material and the accompanying video.

\begin{table*}[t]
	\caption{Quantitative comparisons of our approach with state-of-the-art methods. Our approach achieves the best overall performance. The best result is highlighted in bold. Due to the extremely complex lighting in the Veach scene, the ground-truth images used for training contain significant noise, which negatively affects the training process of all methods.}
    
	\footnotesize 
	\resizebox{\linewidth}{!}{
		\begin{tabular}{@{}l|cccc|cccc|cccc@{}}
			\toprule
			\multirow{2}{*}{Methods}   & \multicolumn{4}{c}{Chair Room} & \multicolumn{4}{|c}{Bed Room} & \multicolumn{4}{|c}{Veach Room} \\
			& PSNR$\uparrow$  & SSIM$\uparrow$ & LPIPS$\downarrow$ & Gaussians$\downarrow$ & PSNR$\uparrow$  & SSIM$\uparrow$  & LPIPS$\downarrow$  & Gaussians$\downarrow$ & PSNR$\uparrow$  & SSIM$\uparrow$ & LPIPS$\downarrow$ &  Gaussians$\downarrow$ \\ \midrule
            NRC                       & 19.27&  0.753&0.348&-     & 21.39& \textbf{0.893}& \textbf{0.082}&-   & \textbf{18.52}& \textbf{0.352}& 0.692&- \\ 
            \midrule
			3DGS                       &24.43 &0.799 &0.283&464253     &11.25 &0.199 &0.697  &3907260   &10.28 &0.106 & 0.632& 1402584\\ 
			SuGaR                      &24.54 &\textbf{0.805}  &0.281&464692     &11.80&0.245&0.710&3676955   &12.24&0.342&0.711&1384693 \\ 
			GaussianMesh                &16.68 &0.590  &0.526&464253     &13.78 &0.324 &0.724  &3907260   &15.73 &0.322 &0.662  &1402584 \\ 
                GaMeS   &19.41 &0.637  &0.447 &479090     &13.27 &0.313 &0.739  &7496705   &16.70 &0.325 &0.642  &1935955 \\ 
			Ours                & \textbf{31.24}& 0.804 &\textbf{0.051} &\textbf{401375}     &\textbf{23.75} &0.766 &0.091  &\textbf{2160865} & 17.12&0.317 &\textbf{0.408}  & \textbf{1365333}\\  \bottomrule
	\end{tabular}}
	\label{table1}
\end{table*}



\subsection{Application of Real-time Shared Rendering}

To further validate the effectiveness of our approach, we extend it to shared rendering, enabling real-time update and rendering for multi-viewer environments.

Unlike the offline training process used in previous experiments, the 3D Gaussian representation is initialized directly from the scene and updated online during rendering. Specifically, sampled viewpoints are rendered using 1-spp path tracing, and the resulting images are used to update the color attributes of the 3D Gaussians in real time. Due to the high-quality initialization provided by our sampling-guided construction, the geometric attributes of the Gaussians remain stable. Therefore, we freeze all Gaussian attributes except for color and update only the color values to cache global illumination efficiently.

We compare our approach with state-of-the-art real-time path tracing and denoising methods, including path tracing (PT), PT+NRC \cite{RN9}, PT+Nvidia OptiX AI-accelerated Denoiser (ONND) \cite{RN68}, and PT+NRC+ONND. The path tracer is configured with 1 sample per pixel (1-spp) and 5 light bounces. Since NRC also performs online optimization using 1-spp path-traced samples, we adopt the same setting to ensure a fair comparison.

We conduct experiments to measure the average performance over a 100-frame sequence in each scene. The quantitative results are shown in Table \ref{table2}. Our approach achieves the best performance in terms of PSNR, SSIM, and Flip compared to all baselines, demonstrating superior perceptual quality. Table \ref{table2} also reports the average rendering time for the 100-frame sequence. Benefiting from efficient Gaussian update and splatting procedures, our method achieves better time efficiency while producing more realistic shadows, textures, and geometric details compared with existing methods. Furthermore, our approach can efficiently cache global illumination and adapt to dynamic lighting conditions. 
\begin{table*}[t]
	\caption{Quantitative comparisons of our approach with state-of-the-art real-time path tracing and denoising methods. Our approach achieves the best overall performance. The best result is highlighted in bold.}
 	\footnotesize 
  \resizebox{\linewidth}{!}{
	\begin{tabular}{@{}l|cccc|cccc|cccc@{}}
		\toprule
		\multirow{2}{*}{Methods}   & \multicolumn{4}{c}{White Room} & \multicolumn{4}{|c}{Living Room} & \multicolumn{4}{|c}{Sun Temple} \\
		& PSNR$\uparrow$  & SSIM$\uparrow$ & Flip$\downarrow$ & Time (ms)$\downarrow$ & PSNR$\uparrow$  & SSIM$\uparrow$  & Flip$\downarrow$ & Time (ms)$\downarrow$ & PSNR$\uparrow$  & SSIM$\uparrow$ & Flip$\downarrow$ & Time (ms)$\downarrow$ \\ \midrule
		PT                         & 13.38 &0.405&0.283&  \textbf{1.4}     & 15.22 & 0.368&0.267&      \textbf{1.2} & 13.61 &0.550&0.182&    \textbf{1.0}  \\ 
		PT+NRC                     & 18.93 &0.522&0.153&  3.1     & 17.94 & 0.433&0.179&      2.4 & 16.20 &0.570&0.143&    2.2   \\
		PT+ONND          & 24.82 &0.848&0.128&  3.0     & 28.32 & 0.897&0.114&      2.5 & 29.53 &0.902&0.087&    2.5   \\
		PT+NRC+ONND      & 25.73 &0.873&0.100&  4.8     & 28.05 & 0.878&0.132&      3.6 & 25.91 &0.885&0.156&    3.8   \\
		Ours                       & \textbf{31.04} &\textbf{0.921}&\textbf{0.063}&  3.3     & \textbf{31.09} & \textbf{0.956}&\textbf{0.049}&      1.7 & \textbf{31.23} &\textbf{0.942}&\textbf{0.042}&    2.8   \\ \bottomrule
	\end{tabular}}
	\label{table2}
\end{table*}


\subsection{Ablation Studies}

In this section, we analyze the effectiveness of the key components in our framework through ablation studies. Specifically, we remove the following components and compare them with the full model: the white-box construction strategy, the albedo–shading decomposition strategy, and the Neural Illumination Enhancement module. All ablation experiments are conducted on three room-scale scenes and three object-level scenes. The quantitative results are reported in Table \ref{table3}. Additional details shown in the supplementary material.

\textit{White-box Construction.}
To evaluate the effectiveness of the white-box construction strategy, we compare it with a baseline that randomly initializes Gaussians within the scene bounding box. As shown in Table \ref{table3}, the proposed white-box construction enables more effective global illumination caching while reducing redundant computation. 

\textit{Albedo–Shading Decomposition Strategy.}
The proposed albedo–shading decomposition strategy reduces the maximum spatial frequency of the light field, allowing larger sampling intervals while maintaining reconstruction quality and better preserving texture details. As shown in Table \ref{table3}, removing this strategy requires the model to directly represent higher-frequency illumination variations and more complex texture patterns, which leads to degraded rendering quality.

\textit{Neural Illumination Enhancement.}
To model complex non-Lambertian effects, we introduce a Neural Illumination Enhancement to capture view-dependent effects such as specular highlights. In the ablation experiment, we remove Neural Illumination Enhancement module and rely solely on explicit 3D Gaussians to represent illumination. The results show that without the Neural Illumination Enhancement module, the model struggles to accurately reproduce specular highlights and other view-dependent reflections, while the full model can better recover these complex lighting effects.


\begin{table*}[t]
	\caption{Ablation studies on the white-box construction, albedo-shading decomposition, and the NIE. The best results are highlighted.}
 	\footnotesize 
  \resizebox{\linewidth}{!}{
	\begin{tabular}{@{}l|ccc|ccc|ccc@{}}
		\toprule
		\multirow{2}{*}{Methods}   & \multicolumn{3}{c}{Bed Room} & \multicolumn{3}{|c}{Chair Room} & \multicolumn{3}{|c}{Veach Room} \\
		& PSNR$\uparrow$  & SSIM$\uparrow$ & LPIPS$\downarrow$  & PSNR$\uparrow$  & SSIM$\uparrow$  & LPIPS$\downarrow$  & PSNR$\uparrow$  & SSIM$\uparrow$  & LPIPS$\downarrow$  \\ \midrule
		w/o white-box cons.                          &21.78  &0.754&0.153     & 30.98& \textbf{0.807}&     0.097& 14.89&\textbf{0.323}&     0.605\\ 
		w/o albedo--shading dec.                    &23.70 &0.770& \textbf{0.086}  & 30.56 &0.799 &0.065&16.17&0.280&0.436\\
		w/o residual network          & 22.81 &0.762&  0.096  &30.95 & 0.785&0.068&\textbf{17.88}&0.292&0.494\\
		Ours                       & \textbf{24.24} &\textbf{0.773}&0.089    & \textbf{31.24} & 0.804&\textbf{0.051} & 17.12&0.317&\textbf{0.408}\\ \bottomrule
	\end{tabular}}
	\label{table3}
    
  \resizebox{\linewidth}{!}{
	\begin{tabular}{@{}l|ccc|ccc|ccc@{}}
		\toprule
		\multirow{2}{*}{Methods}   & \multicolumn{3}{c}{ Testball} & \multicolumn{3}{|c}{Copper Man
} & \multicolumn{3}{|c}{Kettle} \\
		& PSNR$\uparrow$  & SSIM$\uparrow$ & LPIPS$\downarrow$  & PSNR$\uparrow$  & SSIM$\uparrow$ & LPIPS$\downarrow$ 
& PSNR$\uparrow$  & SSIM$\uparrow$ & LPIPS$\downarrow$ \\ \midrule
		w/o white-box cons.                          &27.17&0.908&0.036& 31.04& 0.947&     0.048
& 26.73&0.797&     0.106\\ 
		w/o albedo--shading dec.                    &20.20&0.879& 0.158& 29.85    &0.933&0.063
&19.54&0.751&0.244\\
		w/o residual network          & 26.98&0.907&  0.038&21.55     & 0.877&0.186 
&26.19&0.778&0.103\\
		Ours                       & \textbf{27.18} &\textbf{0.909}&\textbf{0.036}& \textbf{31.10} & \textbf{0.949}&\textbf{0.048}  & \textbf{26.76} &\textbf{0.797}&\textbf{0.103} \\ \bottomrule
	\end{tabular}}
	\label{table3}
\end{table*}
\section{Conclusion}

In this paper, we analyze the reconstruction of 3D Gaussian Splatting (3DGS) from the perspective of light field sampling and propose a novel white-box 3DGS construction framework based on plenoptic sampling theory. The proposed method constructs 3D Gaussians directly from mesh geometry, enabling efficient caching of global illumination with a small number of sampled views. 

Furthermore, we introduce an efficient optimization strategy that combines explicit Gaussian representations with Neural Illumination Enhancement module for modeling complex view-dependent reflections. The explicit Gaussians capture the dominant global illumination, while the Neural Illumination Enhancement module is selectively activated to represent non-Lambertian effects such as specular highlights.

Extensive experiments on a diverse set of scenes demonstrate that our approach achieves superior rendering quality compared with existing Gaussian-based and neural caching methods. In addition, the proposed framework enables efficient real-time updates using low-sample path-traced signals, making it suitable for applications such as real-time shared rendering.
\newline
\indent
\textbf{Limitations and Future Work.} Our method has several limitations. First, the proposed white-box construction relies on the availability of reliable mesh geometry to guide the initialization and spatial distribution of 3D Gaussians. Consequently, the reconstruction quality may be affected when the input mesh is inaccurate or unavailable. While this assumption is well suited for many virtual environments and 3D content assets where geometry is readily available, applying the framework to real-world scenes requires an additional geometry reconstruction step. Future work will explore integrating the proposed framework with more robust multi-view reconstruction techniques.

Second, although the Neural Illumination Enhancement module effectively models view-dependent effects, it is currently disabled in the real-time shared rendering setting to maintain high update rates. As a result, the real-time mode relies solely on explicit Gaussian representations, which may limit the ability to reproduce complex non-Lambertian effects. In future work, we plan to investigate lightweight neural representations or pre-trained neural components that can be efficiently incorporated into dynamic rendering scenarios.

\bibliographystyle{splncs04}
\bibliography{main}
\end{document}